\documentclass{article}

\usepackage[english]{babel}

\usepackage[letterpaper,top=2cm,bottom=2cm,left=3cm,right=3cm,marginparwidth=1.75cm]{geometry}

\usepackage{amsmath}
\usepackage{graphicx}
\usepackage[colorlinks=true, allcolors=blue]{hyperref}

\usepackage{bm}
\usepackage{xcolor}
\usepackage{comment}
\usepackage{hyperref}
\usepackage{caption}
\usepackage{amsmath}
\usepackage{bbold}
\usepackage{etoolbox}
\usepackage{adjustbox}
\usepackage{makecell}
\usepackage{subcaption}
\usepackage{textcomp}
\usepackage[symbol]{footmisc}
\usepackage{booktabs}
\usepackage{multirow}
\usepackage{algorithm}
\usepackage{algorithmic}
 \usepackage{setspace}
 \usepackage{textcomp}
\usepackage[sort&compress,numbers]{natbib}
\providecommand{\keywords}[1]{\textbf{Keywords:} #1}

\title{Chemical Reaction Neural Networks for Fitting Accelerating Rate Calorimetry Data}

\author{\stepcounter{footnote}Saakaar Bhatnagar$^{a}$\thanks{Corresponding Author} \\ {\small \textit{sbhatnagar@altair.com}} \and Andrew Comerford$^{a}$ \\ { \small \textit{acomerford@altair.com}} \and Zelu Xu$^{a}$ \\ { \small \textit{zxu@altair.com}}  \and Davide Berti Polato$^{b}$ \\ { \small \textit{davide.bertipolato@beond.net}} \and Araz Banaeizadeh$^{a}$ \\ { \small \textit{araz@altair.com}}  \and    Alessandro Ferraris$^{b}$ \\ { \small \textit{alessandro.ferraris@beond.net}} \\ }

\date{$^{a}${\small Altair Engineering Inc., 640 W. California Ave, Sunnyvale, CA, USA}\\ $^{b}${\small BeonD Srl - Corso Castelfidardo, 30A, 10129 Torino, Italy}} 

\begin{document}

\maketitle

\keywords{ Thermal Runaway, Accelerating Rate Calorimetry (ARC), Artificial Intelligence (AI), Chemical Reaction Neural Networks (CRNN), Neural ODE     } \\

\begin{abstract} \centering As the demand for lithium-ion batteries rapidly increases there is a need to design these cells in a safe manner to mitigate thermal runaway. Thermal runaway in batteries leads to an uncontrollable temperature rise and potentially battery fires, which is a major safety concern. Typically, when modelling the chemical kinetics of thermal runaway calorimetry data ( e.g. Accelerating Rate Calorimetry (ARC))  is needed to determine the temperature-driven decomposition kinetics. Conventional methods of fitting Arrhenius Ordinary Differential Equation (ODE) thermal runaway models to ARC data make several assumptions that reduce the fidelity and generalizability of the obtained model. In this paper, Chemical Reaction Neural Networks (CRNNs) are trained to fit the kinetic parameters of N-equation Arrhenius ODEs to ARC data obtained from a Molicel 21700 P45B. The models are found to be better approximations of the experimental data. The flexibility of the method is demonstrated by experimenting with two-equation and four-equation models. Thermal runaway simulations are conducted in 3D using the obtained kinetic parameters, showing the applicability of the obtained thermal runaway models to large-scale simulations.
\end{abstract}

\section{Introduction}
\label{sec:intro}

Thermal runaway in battery packs is a major safety concern for commercial applications such as electric vehicles, potentially leading to catastrophic outcomes like battery pack fires. This phenomenon occurs due to thermal abuse conditions that lead to exothermic degradation reactions of battery components, such as anode decomposition, cathode conversion, SEI decomposition, and electrolyte breakdown\cite{feng2018thermal,spotnitz2003abuse}. Typical thermal abuse failure modes include, but are not limited to, physical damage, internal short circuits, overcharging, or overheating (e.g., extreme temperature exposure)\cite{feng2018thermal}. The heat released under such conditions, when a cell or group of cells fails, can lead to a chain reaction where adjacent cells enter a self-heating state and undergo thermal runaway\cite{Feng2016}. This propagation can consume an entire battery module or pack. These safety concerns are even more pressing in today's electrification environment, particularly as the industry moves towards higher power and energy density cells\cite{Golubkov2014,feng2018thermal}. To address these concerns, cell and pack manufacturers must adhere to strict safety protocols to avoid catastrophic outcomes. Simulation-driven design offers a platform to optimize designs and aid in the prevention and mitigation of thermal runaway. For example, thermal analysis of novel heat shield materials can be conducted efficiently to understand their effectiveness at mitigating propagation.\\

Typically, when modeling thermal runaway, chemical decomposition kinetics are represented by multiple Arrhenius Ordinary Differential Equations (ODEs)\cite{Hatchard2001,spotnitz2003abuse,Kim2007}. These reaction rates, combined with the reaction enthalpy, allow for the prediction of chemical heat generation under abuse scenarios. For each of the Arrhenius ODEs, the kinetic parameters and heat of reaction must be determined. In the literature, a popular model is that proposed by \citet{Hatchard2001}. This model utilized calorimetry-derived data to model thermal runaway in an 18650 Lithium Cobalt Oxide (LCO) cell for the anode, cathode, and SEI decomposition reactions. The model utilized a lumped approach, that is treating the jellyroll as a single temperature, and demonstrated excellent agreement with experimental oven tests. This model has been the basis of a number of subsequent models and has been extended to simulate other cell types, 3D domains (i.e. not lumped), additional decomposition reactions (e.g., electrolyte breakdown and short circuit events), and different cathode chemistries\cite{Kim2007,Peng2016,Coman2017, Bugryniec2020,Kong2021}. In addition to these Hatchard-derived models, parameter estimation for the generic Arrhenius ODE have been proposed in the literature. The model by \citet{Shelkea2022}, which is similar to \cite{Hatchard2001} and \cite{Kim2007}, but with the addition of a mix term to match the experimental accelerating rate calorimetry (ARC) data, is one example. \citet{ping2017modelling} demonstrated a five-stage model also based on ARC data. The kinetic and reaction heat parameters were determined through a combination of parameter fitting and literature data. More details regarding this method are provided later. \citet{Feng2018} proposed an elaborate mathematical model based on literature (e.g. \cite{Hatchard2001}) and some novel formulations, demonstrating excellent agreement with the ARC experimental data. Their model utilized parameters from the literature and adjusted the parameters to fit experimental data; limited details of the fitting methods were provided. \citet{chen2021simplified} introduced an ARC-derived two-stage model. The advantage of this model is it requires fewer parameters than the four-stage NREL model. Their approach demonstrated good agreement with the ARC experimental data. In general, the models currently in literature can intricately model the physics of thermal runaway, but parameter identification often relies on literature data or laborious and not necessarily robust identification methods.\\

Parameter estimation for thermal runaway modelling remains a major challenge and typically relies on calorimetry data to measure chemical heat release. In literature, there are several types of calorimetry data used, each with their own advantages and disadvantages. Some examples include: ARC \cite{richard1999accelerating,wu2022dimensionless,lei2017experimental,Feng2018,Shelkea2022, chen2021simplified}; Fractional Thermal Runaway Calorimetry (FTRC) developed by NASA \cite{walker2022effect}; and Differential Scanning Calorimetry (DSC) \cite{Ren2018,spotnitz2003abuse,wang2021thermal}. FTRC can accurately measure heat generation and gas release during thermal runaway, offering the advantage of shorter experimental time frames compared to other calorimetry-based approaches. To the authors’ knowledge, no fitting has been done on this data yet, but it represents a promising method. DSC is a thermal analysis technique that measures heat flow into or out of a sample (e.g. cathode material) as it undergoes a controlled temperature ramp. This heat flow can be used to characterize the heat generation of individual battery components, separately. Finally, ARC is one of the most common ways to characterize battery thermal runaway response for a single cell. In this experiment, the intact cell is first placed inside a thermally isolated cell holder in a sealed calorimeter chamber. The test initiates in a heat-wait-seek mode, where a heater heats the calorimeter incrementally, then pauses to allow thermal transients to decay and equilibrium conditions to be reached. If any increase in temperature rate surpasses a preset threshold value (indicating exothermic reactions), the calorimeter switches to exotherm mode. In this mode, the calorimeter tracks the temperature rise and maintains zero temperature differential between the cell surface and the calorimeter walls, ensuring adiabatic conditions. The final phase involves either full thermal runaway and subsequent cooling or simply cooling if the temperature rate falls below the preset threshold value.\\

Typically in the literature, the most common method for determining reaction heat and kinetic parameters from ARC calorimeter-based data involves introducing temperature stages and linearizing the governing Arrhenius kinetic ODE within these stages\cite{chen2021simplified,ping2017modelling}. Parameters are then determined using least squares fitting. 
Dividing the data into more stages generally leads to a better representation of the underlying data. These types of thermal runaway models can be considered cell-based models as they do not directly elucidate the individual reaction kinetics of the electrodes\cite{wang2024}. Alternative models that provide more detailed reaction kinetic information, are the DSC component-based models (e.g. \cite{Ren2018,koenig2023accommodating}). Unfortunately, for the cell-based models, the numerous approximations made during the linearization process reduce the fidelity of the model and often fail to accurately represent the underlying data (see section \ref{sec:linear_fitting}). Some of these inaccuracies can be alleviated through manual tuning, ultimately yielding a reasonable representation of the original data and the physical phenomena can be obtained, albeit through a laborious process. 

Chemical Reaction Neural Networks (CRNNs) have been proposed to discover reaction coefficients, parameters and pathways in systems governed by the Arrhenius equations \cite{ji2021autonomous}. In CRNNs, the differential equation

\begin{equation}
    \label{eqn:CRNN_basic}
    \frac{d\boldsymbol{c}}{dt}= f(\boldsymbol{c},\boldsymbol{\theta},t),
\end{equation}

is integrated in a fully differentiable manner w.r.t \(\boldsymbol{\theta}\), similar to Neural ODEs \cite{chen2018neural}, where f is the neural network formulated using the system of ODEs governing the chemical reactions modelled, \(\boldsymbol{\theta}\) are time-independent parameters to be learnt using stochastic gradient descent using a loss metric, and \textbf{c} represents the time-dependent system state. 

\citet{puliyanda2023benchmarking} used CRNNs to obtain chemical kinetic models from noisy spectroscopic data, achieving significant performance improvements over least squares regression. \citet{kircher2024global} proposed the Global Reaction Neural Network (GRNN) architecture, embedding thermodynamic and stoichiometric prior knowledge into the network architecture, which autonomously learned kinetic models directly from noisy reactor data. \citet{koenig2023accommodating} used CRNNs to learn the thermal decomposition models of Nickel-Cobalt-Manganese (NMC) cathodes using DSC data, obtaining more physically consistent fits by avoiding empirical assumptions made when fitting using the Kissinger method\cite{kissinger1956variation}. These studies demonstrated that, since the learned parameters are constrained to obey the governing ODEs, the physics-informed nature of the CRNN method outperformed other data-driven methods in terms of better accuracy and lower data requirements. Additionally, the resulting models are physically interpretable in the form of the governing ODEs.\\

This work proposes a novel method to obtain accurate fits for Arrhenius equation-based thermal runaway models from experimental ARC data. CRNNs are utilized to learn the required parameters. A staging and linearization-based fitting method is demonstrated on the data, showing its shortcomings in creating accurate models. The obtained parameters are then tuned via training a CRNN, resulting in learned models that better represent the data. The experiment is conducted on two and four-stage models, showing the flexibility of the method. The tuned parameters are tested in 3D simulations of ARC and oven tests, demonstrating improved workflow and performance compared with more traditional methods cited in the literature.

\section{Experimental Setup }
\label{sec:experimental_setup}
\begin{figure}[h!]
\centering
\begin{subfigure}{0.3\textwidth}
    \includegraphics[width=\textwidth]{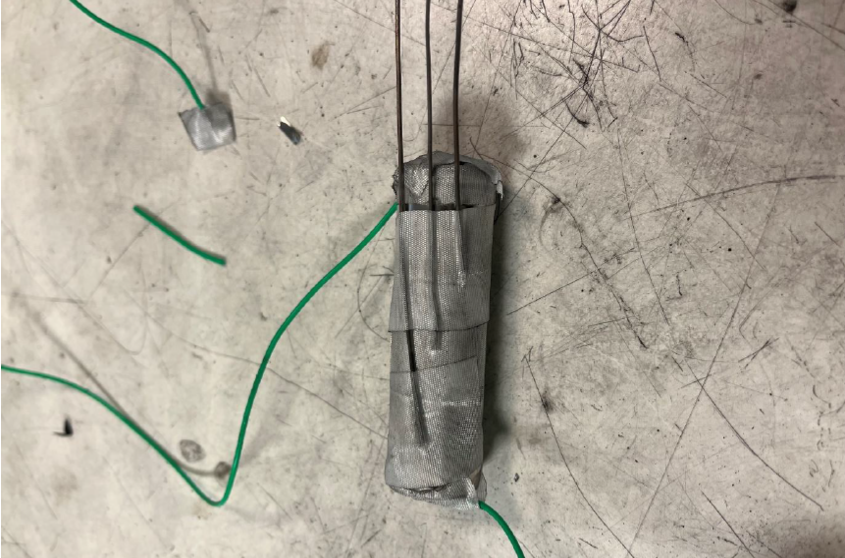}
    \caption{}
    \label{fig:TC}
\end{subfigure}
\begin{subfigure}{0.3\textwidth}
    \includegraphics[width=\textwidth]{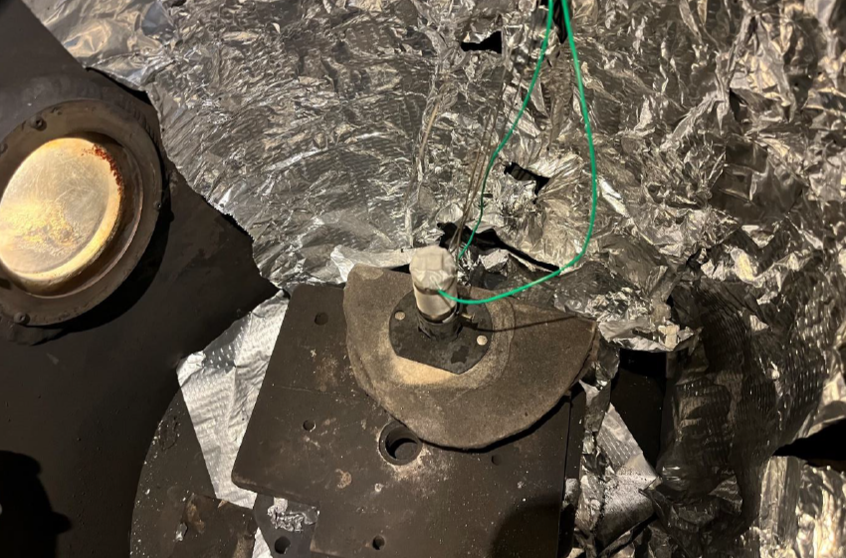}
    \caption{}
    \label{fig:setup}
\end{subfigure}
\begin{subfigure}{0.3\textwidth}
    \includegraphics[width=\textwidth]{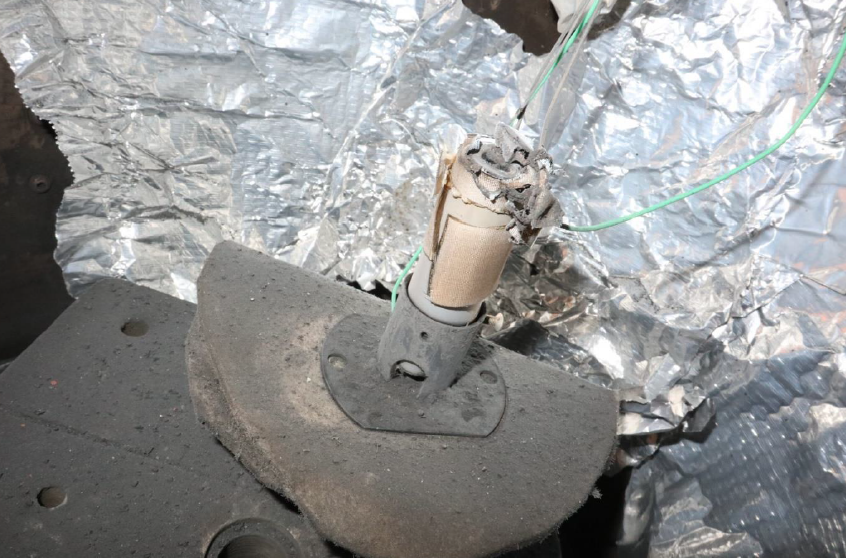}
    \caption{}
    \label{fig:post}
\end{subfigure}
\begin{subfigure}{0.48\textwidth}
    \includegraphics[width=\textwidth]{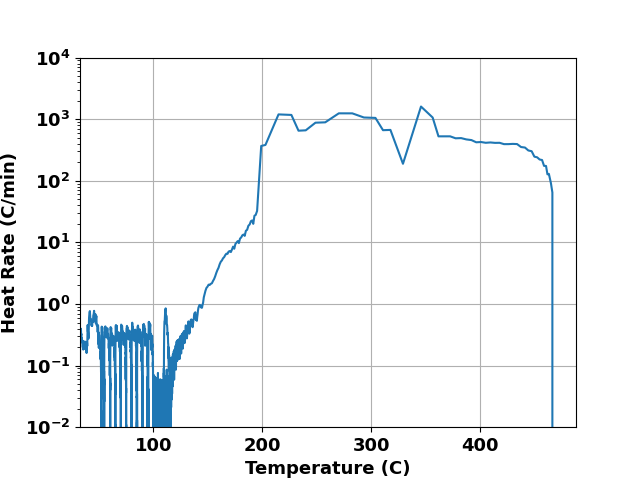}
    \caption{}
    \label{fig:temp_rate_from_exp}
\end{subfigure}
\hfill
\begin{subfigure}{0.48\textwidth}
    \includegraphics[width=\textwidth]{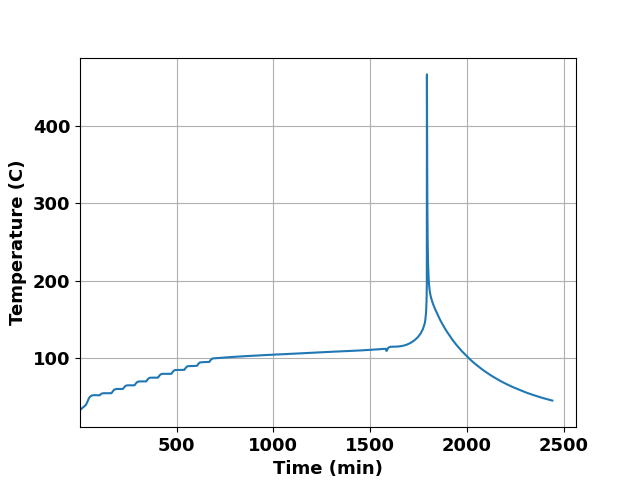}
    \caption{}
    \label{fig:time_temp_from_exp}
\end{subfigure}

\caption{\centering ARC experimental testing (\subref{fig:TC}) Location of  thermocouples on the cell surface to measure cell temperature  (\subref{fig:setup}) Cell in holder inside ARC chamber (\subref{fig:post}) Burnt out cell post ARC test (\subref{fig:temp_rate_from_exp}) Heat rate versus temperature (\subref{fig:time_temp_from_exp}) Temperature evolution with time}
\label{Figure:ARC_experiment}
\end{figure}

The battery cell under investigation was a Molicel 21700 P45B with a mass of 0.066kg and a capacity of 4.5Ah. The sample cell was inserted into a holder within the ARC reactor. The cell was preconditioned by gradually heating it to a controlled temperature of 50$^oC$. These conditions were allowed to equilibrate to ensure a uniform baseline temperature. To investigate thermal runaway, a Heat-Wait-Seek (HWS) testing protocol was employed. Initially, the cell underwent a temperature ramp exceeding 0.02$^oC/min$ to bring it close to the onset of self-heating. Once stabilized at this temperature, the ARC initiated the HWS protocol. During this protocol, the sample was heated in increments of 5 $^oC$ and then allowed to equilibrate (“wait”) for over 40 minutes. Adiabatic conditions were maintained by adjusting the chamber temperature to match that of the cell. Subsequently, the calorimeter entered a “seek” mode for 10 minutes to detect any exothermic activity: If any signs of self-heating (e.g., a sustained temperature rise of 0.02 $^oC/min$ ), the ARC transitioned to exotherm mode, otherwise the heat-wait-seek cycle was repeated. In exotherm mode, the ARC tracked the temperature rise due to exothermic reactions and endeavored to maintain adiabatic conditions by dynamically adjusting the chamber temperature and provide detailed thermal behavior such as cell (top, middle and bottom) as well as ambient temperature. The setup of the cell within the chamber and the location of thermocouples on the cell surface are shown in Figures (\ref{fig:TC})-(\ref{fig:post}).

The collected ARC data is used to determine the heat rate (in \(^{o}C/min\)) versus temperature (in \(^{o}C\)), as well as temperature versus time. Figures (\ref{fig:temp_rate_from_exp}) and (\ref{fig:time_temp_from_exp}) show the collected data for heat rate versus temperature and temperature versus time, respectively.\\

\section{Methods and Models}

\subsection{TR ODE System}
\label{sec:TR_ODE_sys}

The thermal runaway modelling equations in this work take the form of an Arrhenius reaction equation to model the decomposition kinetics and are given by,

\begin{equation}
    \label{eqn:arrhenius}
    \frac{dc_{i}}{dt}= f_{i}(c_{i})A_{i}e^{\frac{-E_{a,i}}{k_{b} T}}, 
    \textit{       i=1,2...,N} 
\end{equation}

\begin{equation}
    \label{eqn:conc_form}
    f_{i}(c_{i})= c_{i}^{n_{i}}(1-c_{i})^{m_{i}},
\end{equation}

where \(c_{i}\) is the normalized reactant concentration, \(A_{i}\) is the frequency factor, \(E_{a,i}\) is the activation energy, \(k_{b}\) is the Boltzmann constant, T is the temperature, and N refers to the number of stages in the model. \(m_{i}\) and \(n_{i}\) are reaction orders from the rate law. The reaction orders determine whether the reaction is nth order (when $m_{i}=0$) or an auto-catalytic type i.e. the reaction increases as the product is generated ($m_{i}>0$) \cite{brown1997prout,grandjacques2021thermal}.

The heat \(Q_{i}\) from each exothermic stage decomposition can be calculated using

\begin{equation}
    \label{eqn:heat_arrhenius}
    \dot{Q_{i}}=h_{i}\frac{dc_{i}}{dt},
\end{equation}

where \(h_{i}\) is the reaction enthalpy. The temperature update of the cell due to the thermal runaway can then be computed using

\begin{equation}
    \label{eqn:arrhenius_temp_update}
    m_{cell}c_{p}\frac{dT}{dt}=\sum_{i=1}^{N}\dot{Q_{i}} + \dot{Q}_{diss},
\end{equation}

where $m_{cell}$ and \(c_{p}\) represent the mass and specific heat of the cell respectively, and \(\dot{Q_{diss}}\) is the dissipative heat flux due to natural convection and radiation. This term is modelled as 

\begin{equation}
    \label{eqn:Q_diss}
    \dot{Q}_{diss}= A_{cell}\bm{[}h_{conv}(T_{\infty}-T)+\epsilon\sigma(T_{\infty}^{4}-T^{4})\bm{]},
\end{equation}

where \(A_{cell}\) refers to the area of the cell, \(h_{conv}\) is the convective heat transfer coefficient \cite{chen2006thermal}, \(\epsilon=0.8\), \(\sigma\) is the Stefan-Boltzmann constant, and \(T_{\infty}\) is the far field temperature.

A good model representing the the thermal runaway process is achieved by obtaining the values of the following kinetic and reaction heat parameters,

  \begin{equation}
        \label{eqn:trainable_params}
   \boldsymbol{\theta_{i}}=\boldsymbol{[}A,E_{a},h,m,n\boldsymbol{]}_{i}, \forall \text{i=1,2..N},
  \end{equation}
  such that the temperature and heat rate profiles (shown in Figure \ref{fig:data_staged_division}) predicted by integrating the model are a close match to the experimental ARC data.

\subsection{Linearized Arrhenius kinetics}
\label{sec:linear_fitting}

The most common method for fitting the unknown parameters described by Equation \ref{eqn:trainable_params} to ARC data involves staging and linearizing the governing equations of the model. The number of stages, N , depends on the specific model. For example, \citet{chen2021simplified} used a two-stage model, whereas \citet{sun2023thermal} and \citet{ping2017modelling} used four and five-stage models, respectively, to describe the thermal runaway process.

\begin{figure}[h!]
\centering
\begin{subfigure}{0.48\textwidth}
    \includegraphics[width=\textwidth]{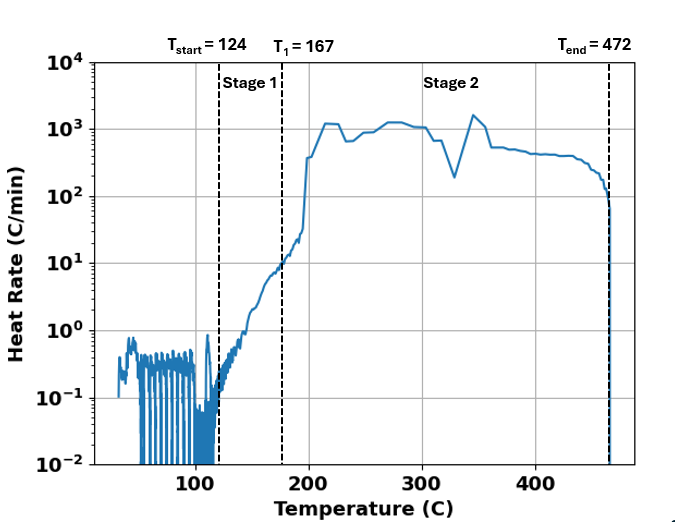}
    \caption{}
    \label{fig:temp_rate_display}
\end{subfigure}
\hfill
\begin{subfigure}{0.48\textwidth}
    \includegraphics[width=\textwidth]{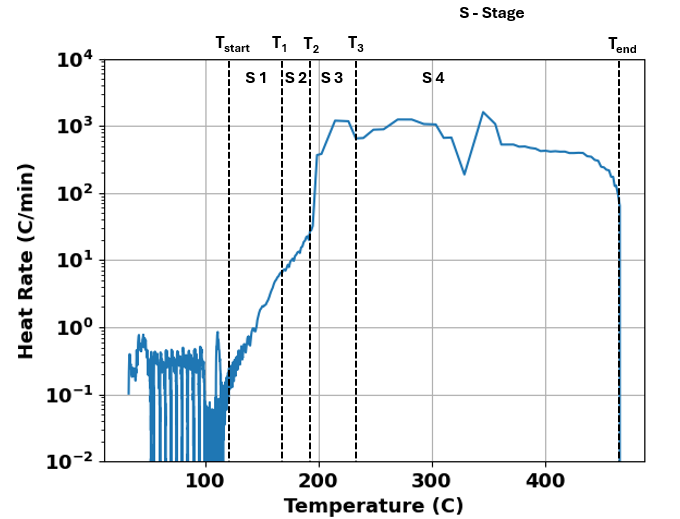}
    \caption{}
    \label{fig:stage_division_4_stage}
\end{subfigure}

\caption{\centering  Division of data into stages, showing the start and end temperature for each stage (\ref{fig:temp_rate_display}) A two-stage fit (\ref{fig:stage_division_4_stage}) A four-stage fit }
\label{fig:data_staged_division}
\end{figure}

Figure (\ref{fig:temp_rate_display}) shows the division of ARC experimental data into two stages, and Figure (\ref{fig:stage_division_4_stage}) shows a four-stage division. In this section, linearization is demonstrated for a two-stage fit. Stage one starts at \(T_{start}\) and ends at \(T_{1}\), and Stage two starts at \(T_{1}\) and ends at \(T_{end}\). Every equation of the model is assigned a stage. The enthalpy of stage \textit{i} is approximated by

\begin{equation}
    \label{eqn:enth_appx}
    h_{i} = mc_{p}(T^{end}_{i}-T^{start}_{i}).
\end{equation}

Equations \ref{eqn:arrhenius} -\ref{eqn:arrhenius_temp_update} are rewritten, through a series of simplifying assumptions described in \cite{chen2021simplified,sun2023thermal} as

\begin{equation}
    \label{eqn:linearized_fit}
    ln \left[ \frac{dT}{dt} \right] =ln \bm{[} A_{i}(T^{end}_{i}-T^{start}_{i}) \bm{]} - \frac{E_{a,i}}{RT}.
\end{equation}

Utilizing the staged experimental data, the slope and intercept of the \(ln(\frac{dT}{dt})\) v/s \(\frac{1}{T}\) plot return the value of \(E_{a,i}\) and \(A_{i}\), respectively.

\begin{figure}[h!]
\centering
\begin{subfigure}{0.48\textwidth}
    \includegraphics[width=\textwidth]{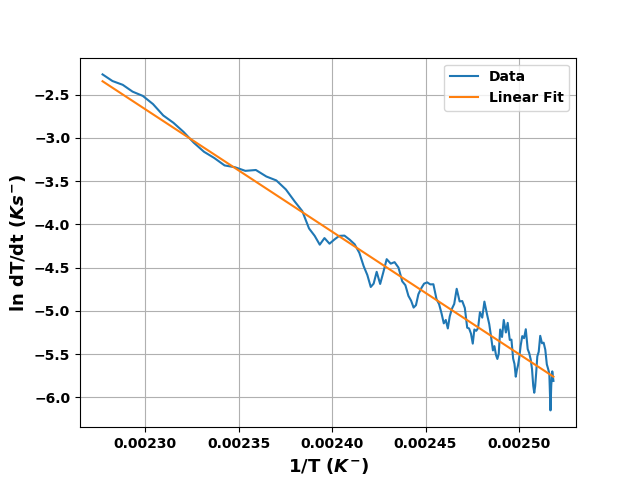}
    \caption{}
    \label{fig:stage_1_linear_fit_display}
\end{subfigure}
\hfill
\begin{subfigure}{0.48\textwidth}
    \includegraphics[width=\textwidth]{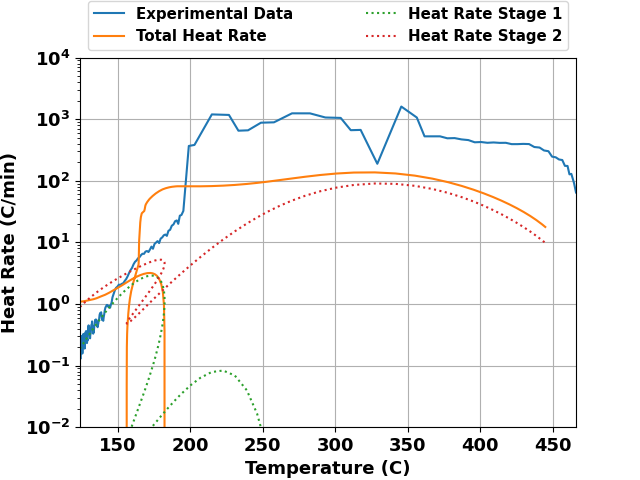}
    \caption{}
    \label{fig:temp_rate_2_stage_linearization}
\end{subfigure}

\caption{\centering (\subref{fig:stage_1_linear_fit_display}) Linear fit of stage 1 (\subref{fig:temp_rate_2_stage_linearization}) Two-stage fit from linearization method }
\label{fig:2_stage_linearization_result}
\end{figure}

The linearization method, although straightforward to implement, relies on a set of assumptions that limits its applicability. For example, it typically assumes a linear decrease in reactant concentration as a function of temperature \cite{sun2023thermal}. However, in practice, integrating the models usually reveals a nonlinear dependency of \(c_{i}\) on T \cite{chen2021simplified}. Furthermore, the reaction orders \((m_{i},n_{i})\) are often adopted from previous studies and, therefore, may not fit the current dataset, resulting in a sub-optimal fit. Finally, the enthalpy condition in Equation \ref{eqn:enth_appx} assumes complete reactant concentration conversion at the end temperature of the stage, which may not hold true upon integrating the ODE models.\\

These assumptions can result in poor fits to experimental ARC data. Figure (\ref{fig:stage_1_linear_fit_display}) shows the linear fit for stage one, which appears to represent the data well. However, Figure (\ref{fig:temp_rate_2_stage_linearization}) illustrates the resultant heat rate of the fit. Unfortunately, the fit does not effectively represent the experimental data.

\subsection{CRNN for ARC data fitting}
\label{sec:CRNN}

\begin{figure}[h!]
\centering

\begin{subfigure}{0.8\textwidth}
    \includegraphics[width=\textwidth]{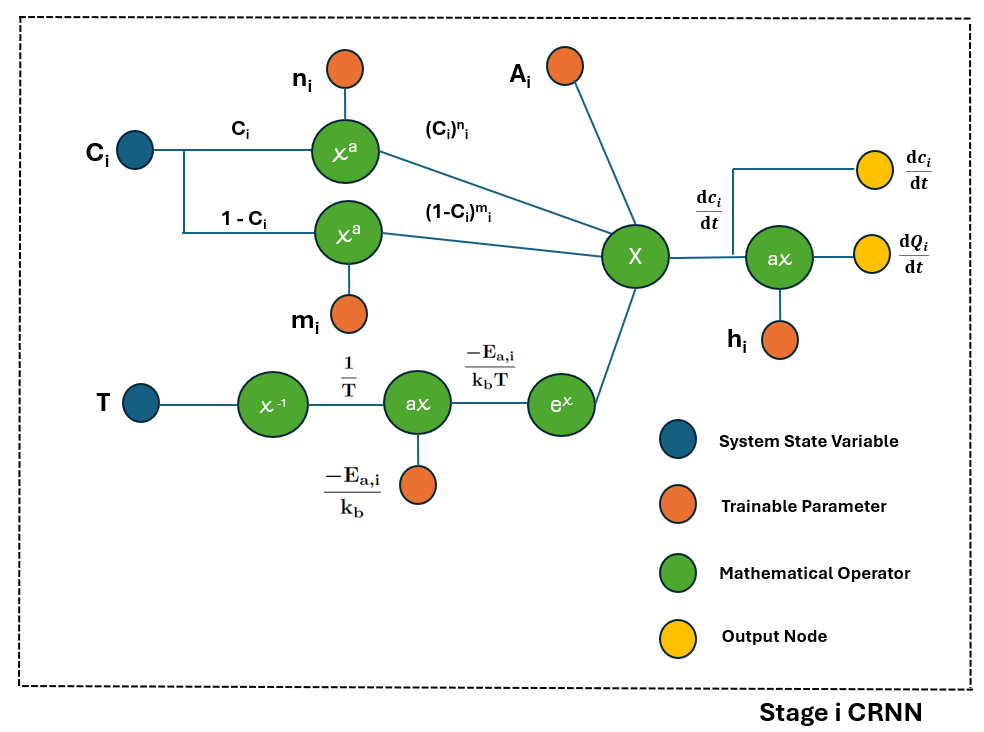}
    \caption{}
    \label{fig:CRNN_stage_i}
\end{subfigure}
\hfill

\begin{subfigure}{\textwidth}
    \includegraphics[width=\textwidth]{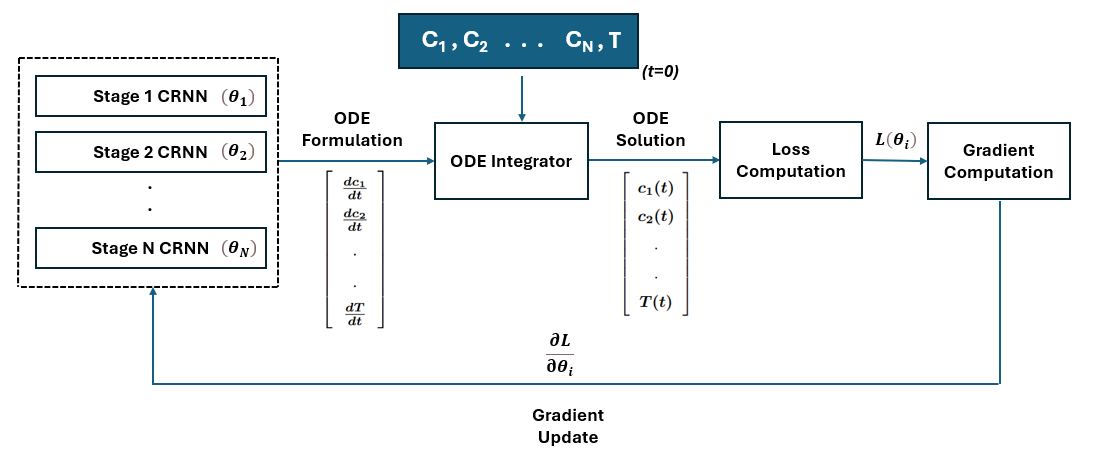}
    \caption{}
    \label{fig:CRNN_workflow_sub}
\end{subfigure}
\hfill

\caption{\centering (\subref{fig:CRNN_stage_i}) CRNN subnetwork diagram for a single stage. The orange nodes denote trainable variables updated via gradient-based optimization, given by \(\boldsymbol{\theta_{i}}=[A_{i},E_{a,i},h_{i},m_{i},n_{i}]\). The blue nodes represent system state variables integrated via a differentiable ODE integrator.  (\subref{fig:CRNN_workflow_sub}) Computation graph for a single CRNN training step. Every CRNN subnetwork (Figure \ref{fig:CRNN_stage_i}) is assimilated to set up the ODE system, which is embedded into the training loop to learn \(\boldsymbol{\theta_{i}}\).}
\label{fig:CRNN_workflow}
\end{figure}

In a CRNN, the parameters to be fitted are treated as trainable variables. Instead of linearizing the governing ODEs, and obtaining the parameters by minimizing a least squares error, the governing ODEs are directly learned using the data, with parameter updates made iteratively via a gradient-based approach. An advantage of this method is that no simplifying assumptions are made on the ODE system, leading to no loss of accuracy or generality during the fitting process. Figure (\ref{fig:CRNN_stage_i}) shows the setup of a CRNN subnetwork for stage \textit{i}, where each stage has a corresponding subnetwork with unique trainable parameters. The orange nodes denote trainable parameters (obtained via gradient-based optimization), and the blue nodes denote state variables, the time histories of which are obtained via ODE integration. The output of a subnetwork is the concentration reaction rate \(\frac{dc_{i}}{dt}\) and the rate of heat generated \(\dot{Q_{i}}\) by stage.  Figure (\ref{fig:CRNN_workflow_sub}) shows how each subnetwork is assimilated into the overall training loop. The resultant system of ODEs is integrated via a differentiable ODE integrator, and the system state \( \boldsymbol{[} c_{1}(t), c_{2}(t), ... c_{N}(t), T \boldsymbol{]}\) is used to compute the loss \textbf{\textit{L}} relative to the experimental data. Reverse-mode automatic differentiation \cite{baydin2018automatic} is leveraged to compute \(\frac{\partial \boldsymbol{L}}{\partial \boldsymbol{\theta_{i}}}\), which is then used to update the values of \(\boldsymbol{\theta_{i}}\) via gradient descent.\\

In this work, the trainable parameters \(\boldsymbol{\theta_{i}}\) in the CRNN are initialized using the linearization method (Section \ref{sec:linear_fitting}). This improved initial guess, over random initialization, accelerates the optimization process. The loss function used to measure the difference between the predicted solution is the \(L_{2}\) norm of temperature history:

\begin{equation}
    \label{eqn:loss_CRNN}
    Loss= \frac{\sum_{j=1}^{N_{ts}}(T^{pred}_{j}-T^{truth}_{j})^2}{N_{ts}},
\end{equation}

where \(N_{ts}\) is the number of timesteps.

The fully differentiable computational training loop was set up using the JAX \cite{jax2018github} automatic differentiation framework. The differentiable ODE solver library Diffrax \cite{kidger2021on}, based on JAX, was used to solve the CRNN ODE system. Kvaerno's \(5^{th}\) order method \cite{kvaerno2004singly} is employed for the numerical stiff ODE solve, and the Adam \cite{kingma2014adam} optimizer is utilized to minimize the loss.  For all results shown, the optimization is done for 10,000 steps with a scheduled learning rate, starting at \(10^{-3}\) and decreasing by a factor of 0.9 every 300 steps.

\section{Results and Discussion}

\begin{table}[]
    \centering
    \
\begin{tabular}{cccc} \toprule
    Name & Mass (Kg) & Area (\(m^{2}\)) & Specific Heat (\(J Kg^{-} K^{-}\)) \\ \midrule
    
    Value  & 0.066 & 4.618E-3 & 859 \\ \bottomrule
\end{tabular}
\caption{Physical properties of the cell}
    \label{tab:cell_params}
\end{table}

A CRNN is used to fit the ARC data shown in Figure \ref{fig:data_staged_division}. The data is divided into N stages, every stage is linearized and a fit for the kinetic parameters \(\boldsymbol{\theta_{i}}\) is obtained \(\forall i=1.....N\). The obtained parameters are used to initialize the CRNN and are improved via optimizing the loss of the CRNN (given by Equation \ref{eqn:loss_CRNN}) iteratively. The fitting process is demonstrated for N=2 and N=4, showing the flexibility of the proposed method. The model obtained from the four-stage fit is used in a 3D simulation of ARC and oven tests, to demonstrate the portability of the resulting thermal runaway models to accurate 3D simulations. 

Table \ref{tab:cell_params} lists the values of the cell properties. The initial values for reaction orders \(m_{i},n_{i}\) are chosen based on commonly observed values in literature.

\subsection{Two-stage model} 
\label{sec:2_stage_results}

A two-stage model of the ARC data takes the form:

    \begin{equation}
        \label{eqn:2_stage_model_1}
        \frac{dc_{1}}{dt}=c_{1}A_{1}e^{(\frac{-E_{a,1}}{RT})},
    \end{equation}

\begin{equation}
    \label{eqn:2_stage_model_2}
    \frac{dc_{2}}{dt}=(1-c_{2})^{m_{2}}A_{2}e^{(\frac{-E_{a,2}}{RT})},
\end{equation}

\begin{equation}
    \label{eqn:2_stage_model_energy}
    mc_{p}\frac{dT}{dt}= \sum_{i=1}^{2}h_{i}\frac{dc_{i}}{dt},
\end{equation}

with Equations \ref{eqn:2_stage_model_1} and \ref{eqn:2_stage_model_2} being particular instances of Equation \ref{eqn:arrhenius}. This model lumps the overall heat generation process into two terms and can still be motivated by the chemical kinetics of the decomposition reactions. For example, during stage one, two dominant decomposition exothermic reactions contribute to heat generation:  the decomposition of the SEI and the reaction of intercalated lithium at the anode with electrolyte. In the second stage, which is the dominant source of heat, the following processes can be observed: electrolyte decomposition; cathode material decomposition and release of oxygen; the reaction of released oxygen with the electrolyte; and potentially an internal short circuit (ISC) due to separator melting. Examples of the different reaction processes can be found in the literature, e.g.  \citet{chen2021simplified,Kim2007,Feng2018,Coman2017} among others. \\

\begin{figure}[h!]
\centering
\begin{subfigure}{0.48\textwidth}
    \includegraphics[width=\textwidth]{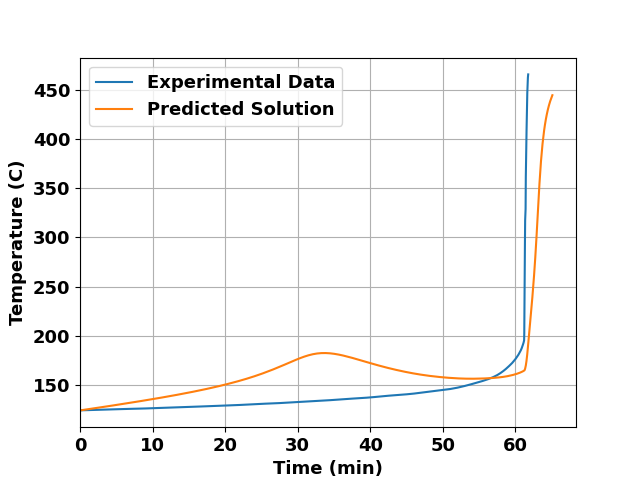}
    \caption{}
    \label{fig:temp_before_2_stage}
\end{subfigure}
\hfill
\begin{subfigure}{0.48\textwidth}
    \includegraphics[width=\textwidth]{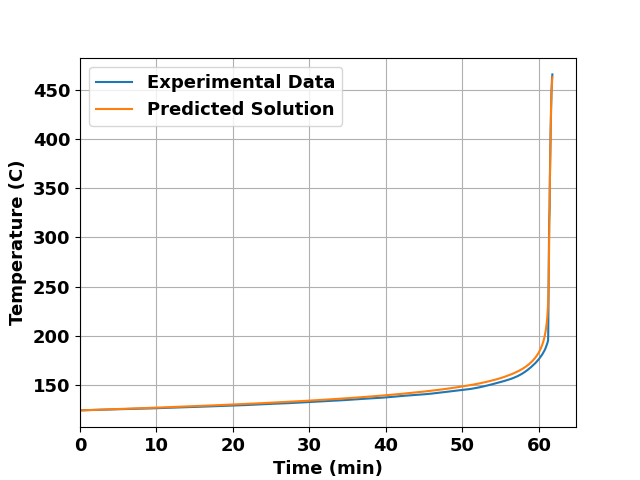}
    \caption{}
    \label{fig:temp_after_2_stage}
\end{subfigure}
\hfill
\begin{subfigure}{0.48\textwidth}
    \includegraphics[width=\textwidth]{pictures/check_rate_before_2_stage.png}
    \caption{}
    \label{fig:rate_before_2_stage}
\end{subfigure}
\hfill
\begin{subfigure}{0.48\textwidth}
    \includegraphics[width=\textwidth]{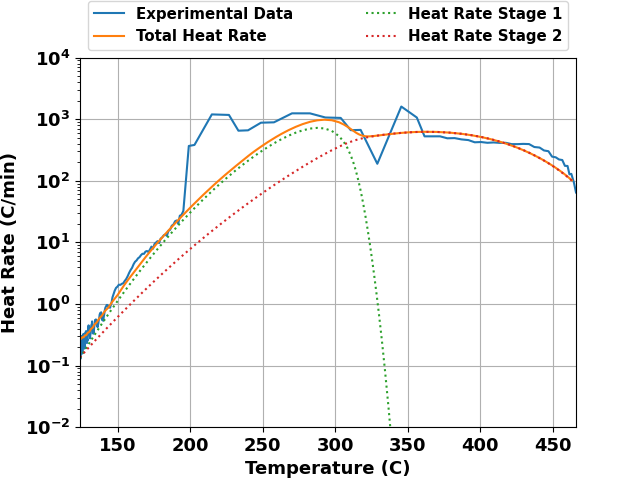}
    \caption{}
    \label{fig:rate_after_2_stage}
\end{subfigure}
\hfill
\caption{\centering (\subref{fig:temp_before_2_stage}) Temperature evolution from linear fitting 
(\subref{fig:temp_after_2_stage}) Temperature evolution after CRNN training 
(\subref{fig:rate_before_2_stage}) Heat rate evolution from linear fitting
(\subref{fig:rate_after_2_stage}) Heat rate evolution after CRNN training
}
\label{fig:2_stage_results}
\end{figure}

The description of the staging and linearization is identical to Section \ref{sec:linear_fitting} and is omitted here for brevity. The temperature values used for the staging, as shown in Figure (\ref{fig:data_staged_division}), were T$_{start}=124^oC$, T$_{1}=167^oC$, and T$_{end}=472^oC$. Figure (\ref{fig:2_stage_results}) shows the integrated solution for parameters initially returned from the linear fit (Figs. (\ref{fig:temp_before_2_stage}),(\ref{fig:rate_before_2_stage})), and for parameters after training a CRNN (Figs. (\ref{fig:temp_after_2_stage}), (\ref{fig:rate_after_2_stage}) ). Figure (\ref{fig:temp_before_2_stage}) shows the temperature increase with time compared to the experimental data, and it is observed that there is a significant difference in the profile of temperature evolution. An overprediction of temperature is observed until the system is close to thermal runaway, where it turns to an underprediction and the thermal runaway time is not correctly predicted. Figure (\ref{fig:rate_before_2_stage}) shows the heat rate plotted against temperature, and a significant deviation of the prediction is observed with respect to the experimental data. 

After CRNN training, where the linear fit results are used to initialize the CRNN parameters, Figures (\ref{fig:temp_after_2_stage}) and (\ref{fig:rate_after_2_stage}) show much better adherence to the experimental data. The temperature leading up to and during thermal runaway is in better agreement with the experimental data, and the thermal runaway time and peak temperature is better predicted. The heat rate prediction is closer to experimental observation as well. The "linear" part of the heat rate increase is better approximated, and the region of large heat rate is approximated by the model more closely, leading to a better fit overall.

Table \ref{tab:2_stage_table} shows each parameter fitted, for each stage, before and after CRNN training. The improved parameters are listed in bold font.

\begin{table}[!h]
    \centering
    \
\begin{tabular}{cccccccc} \toprule
    \textbf{Stage No.} & \textbf{Fit Method} & \textbf{I.C} \((c_{i,0}\)) & \(\boldsymbol{A_{i}} (s^{-})\)  &  \(\boldsymbol{E_{a,i}}\) (J) & \(\boldsymbol{h_{i}}\) (J) & \(\boldsymbol{m_{i}}\) & \(\boldsymbol{n_{i}}\) \\ \midrule
    
    \multirow{2}{*}{\textbf{Stage 1}}  & Linear & 1.0 & 2.1755E11 & 1.9530E-                                  19 & 2434 & 0 & 1  \\ 
                             & CRNN & 1.0 & \textbf{1.723E11} & \textbf{2.027E-19} & \textbf{8336} & 0 & 1  \\ \bottomrule
     \multirow{2}{*}{\textbf{Stage 2}} & Linear & 0.04 & 2.927E7 & 1.474E-19                             & 16963 & 5 & 0  \\ 
                                & CRNN & 0.04 & \textbf{1.994E7} & \textbf{1.554E-19} & \textbf{15970} & \textbf{4.62} & 0  \\ \bottomrule
\end{tabular}
\caption{\centering Parameter values for the two-stage model. For each stage, the top row lists the parameters obtained from linearization, and the boldfaced values in the bottom row are the parameters obtained after CRNN training.}
    \label{tab:2_stage_table}
\end{table}

\subsection{Four-stage model}
\label{sec:4_stage_results}
A four-stage model fitted to the ARC data gives more flexibility and will yield a better approximation of the heat rate data. The model takes the form:

\begin{equation}
        \label{eqn:4_stage_model_1}
        \frac{dc_{1}}{dt}=c_{1}A_{1}e^{(\frac{-E_{a,1}}{RT})},
    \end{equation}

\begin{equation}
    \label{eqn:4_stage_model_2}
    \frac{dc_{2}}{dt}=c_{2}A_{2}e^{(\frac{-E_{a,2}}{RT})},
\end{equation}

\begin{equation}
    \label{eqn:4_stage_model_3}
    \frac{dc_{3}}{dt}=c_{3}^{n_{3}}(1-c_{3})^{m_{3}}A_{3}e^{(\frac{-E_{a,3}}{RT})},
\end{equation}

\begin{equation}
    \label{eqn:4_stage_model_4}
    \frac{dc_{4}}{dt}=c_{4}(1-c_{4})^{m_{4}}A_{4}e^{(\frac{-E_{a,4}}{RT})},
\end{equation}

\begin{equation}
    \label{eqn:4_stage_model_energy}
    mc_{p}\frac{dT}{dt}= \sum_{i=1}^{4} h_{i} \frac{dc_{i}}{dt}.
\end{equation}

The reaction kinetics in this model are more representative of the physical processes occurring during thermal runaway. Figure (\ref{fig:stage_division_4_stage}) graphically shows the division of stages, which were subsequently linearized and the parameters \(\boldsymbol{\theta}_{i}\) obtained. Table \ref{tab:staging_temperatures_4_stage} details the values of the temperatures at which the stages are divided. 

\begin{table}[h!]
    \centering
    \
\begin{tabular}{ccccc} \toprule
    \(T_{start}\) & \(T_{1}\) & \(T_{2}\) & \(T_{3}\) & \(T_{end}\) \\ \midrule
    
     124 & 161 & 191 & 257 & 472 \\ \bottomrule
\end{tabular}
\caption{Staging temperatures for the four-stage fit depicted in Figure (\ref{fig:stage_division_4_stage}), all values in \(^o\)C}
    \label{tab:staging_temperatures_4_stage}
\end{table}

\begin{figure}[h!]
\centering
\begin{subfigure}{0.48\textwidth}
    \includegraphics[width=\textwidth]{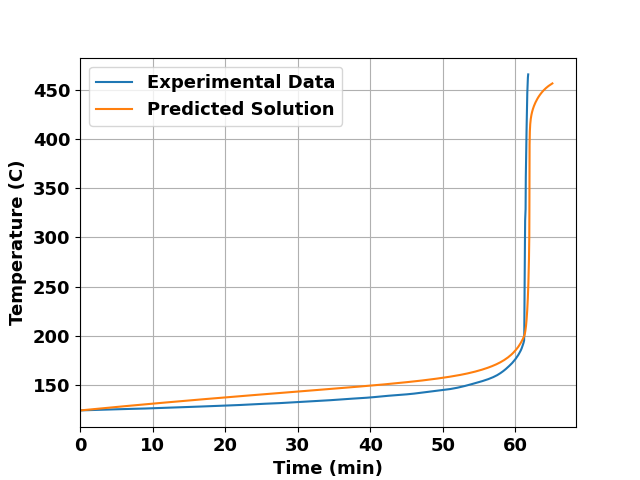}
    \caption{}
    \label{fig:temp_before_4_stage}
\end{subfigure}
\hfill
\begin{subfigure}{0.48\textwidth}
    \includegraphics[width=\textwidth]{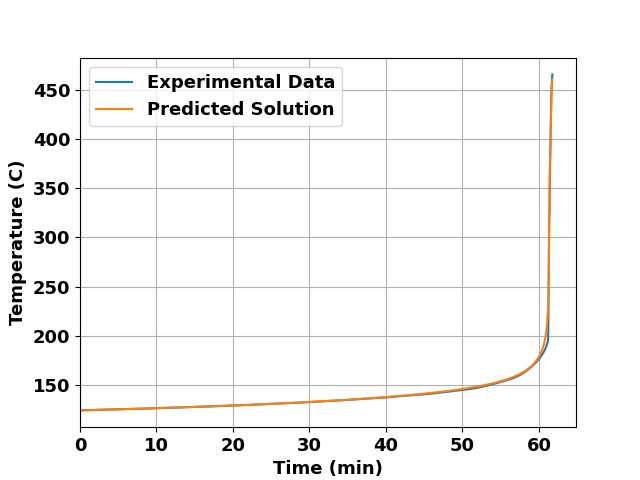}
    \caption{}
    \label{fig:temp_after_4_stage}
\end{subfigure}
\hfill
\begin{subfigure}{0.48\textwidth}
    \includegraphics[width=\textwidth]{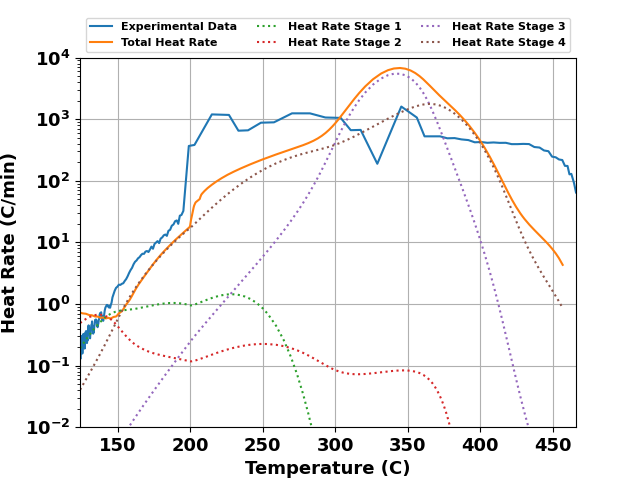}
    \caption{}
    \label{fig:rate_before_4_stage}
\end{subfigure}
\hfill
\begin{subfigure}{0.48\textwidth}
    \includegraphics[width=\textwidth]{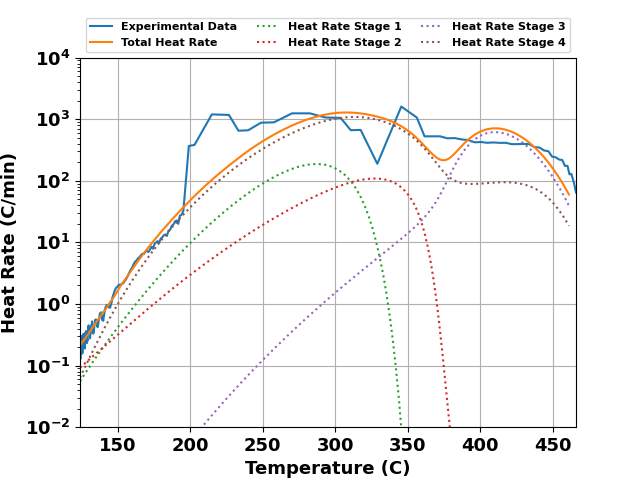}
    \caption{}
    \label{fig:rate_after_4_stage}
\end{subfigure}
\hfill
\caption{\centering (\subref{fig:temp_before_4_stage}) Temperature evolution from linear fitting
(\subref{fig:temp_after_4_stage}) Temperature evolution after CRNN training 
(\subref{fig:rate_before_4_stage}) Heat rate evolution from linear fitting
(\subref{fig:rate_after_4_stage}) Heat rate evolution after CRNN training
}
\label{fig:4_stage_results}
\end{figure}

Similar to the results in Section \ref{sec:2_stage_results}, Figure (\ref{fig:temp_before_4_stage}) shows that the temperature tracked higher relative to the experimental data and was generally a poor representation of the experimental data. Figure (\ref{fig:rate_before_4_stage}) shows the corresponding heat rate; significant differences were observed from the experimental data.

The parameters obtained for the linear fit were used to initialize the CRNN training process. Figure (\ref{fig:temp_after_4_stage}) shows the temperature evolution with time. The fit qualitatively shows good agreement with the experimental data, both in the self-heating phase and during the rapid temperature rise of thermal runaway. Additionally, the fitted data shows a better prediction of the time the system went into thermal runaway and the overall peak temperature. As an additional observation, the obtained temperature profile for the four-stage fit (Figure \ref{fig:temp_after_4_stage}) compares better to the experimental data than with the two-stage fit (Figure \ref{fig:temp_after_2_stage}). This is attributed to the four stages giving the model more ability to accurately represent the data.

Comparing Figure (\ref{fig:rate_before_4_stage}) and (\ref{fig:rate_after_4_stage}), the linear part of the rate curve, corresponding to the slow increase in temperature, is once again better approximated after the CRNN training. The sudden increase in heat rate occurring around 200 \(^{o}\)C is captured slightly more accurately when compared to the two-stage model, leading to a better prediction of the time when thermal runaway occurs.

Table \ref{tab:4_stage_table} shows the parameters for each stage, before and after CRNN optimization. The initial values of \(A_{i}\) and \(E_{a,i}\) for stage 4 are set as the initial values for stage 3. This was done because the parameters obtained via linearization for stage 4 were too poor to use as initial conditions for the CRNN training.

\begin{table}[]
    \centering
    \
\begin{tabular}{cccccccc} \toprule
    \textbf{Stage No.} & \textbf{Fit Method} & \textbf{I.C} \((c_{i,0}\)) & \(\boldsymbol{A_{i}} (s^{-})\)  &  \(\boldsymbol{E_{a,i}}\) (J) & \(\boldsymbol{h_{i}}\) (J) & \(\boldsymbol{m_{i}}\) & \(\boldsymbol{n_{i}}\) \\ \midrule
    
    \multirow{2}{*}{\textbf{Stage 1}}  & Linear & 1.0 & 1.3859E11 & 1.9209E-19 & 2150 & 0 & 1  \\ 
                             & CRNN & 1.0 & \textbf{9.480E10} & \textbf{1.969E-19} & \textbf{2212} & 0 & 1  \\ \bottomrule
     \multirow{2}{*}{\textbf{Stage 2}} & Linear & 1.0 & 4.620E7 & 1.4174E-19                             & 1700 & 0 & 1  \\ 
                                & CRNN & 1.0 & \textbf{2.550E7} & \textbf{1.462E-19} & \textbf{1330} & 0 & 1 \\ \bottomrule

    \multirow{2}{*}{\textbf{Stage 3}} & Linear & 0.04 & 2.371E11 & 1.941E-19                          & 3685 & 2 & 2  \\ 
                                & CRNN & 0.04 & \textbf{3.936E10} & \textbf{2.072E-19} & \textbf{5696} & \textbf{6.34} & \textbf{1.94}  \\ \bottomrule

     \multirow{2}{*}{\textbf{Stage 4}} & Linear & 0.04 & 2.371E11* & 1.941E-19 *                         & 11861 & 5 & 1  \\ 
                                & CRNN & 0.04 & \textbf{2.831E11} & \textbf{1.931E-19} & \textbf{12980} & \textbf{4.61} & 1  \\ \bottomrule
    
\end{tabular}
\caption{\centering Parameters fit for the four-stage model. For each stage, the top row is the parameters obtained from linearization, and the boldfaced values in the bottom row are obtained after CRNN training. The initial values for \(A_{i}\) and \(E_{a,i}\) for stage 4, denoted by *, were initialized to be the same as the initial values for stage 3}
    \label{tab:4_stage_table}
\end{table}

The results obtained in Sections \ref{sec:2_stage_results}  and  \ref{sec:4_stage_results} demonstrate the improvement in parameters achieved through linearization using the CRNN method. The works by  \citet{chen2021simplified}  and \citet{ping2017modelling} utilize the linearization method, which, as shown in previous sections, can result in poor fits. The obtained parameters often require manual fine-tuning, making reproducibility of the method problematic. The CRNN addresses this issue by automatically determining the parameters that best model the experimental data using gradient descent, thus eliminating the need for manual tuning and ensuring reproducibility.

Furthermore, as noted in Table \ref{tab:4_stage_table}, the linear fit for the final stage in a four-stage model can be so poor that some parameters from stage 3 need to be used as initial guesses.  \citet{sun2023thermal} and \citet{feng2018coupled} instead modeled the heat release of the final stage using a cumulative enthalpy method, which measures the total energy released by the last stage of the process and reuses the same function in future simulations. While this method is simple to implement, it does not generalize well to simulations outside of the ARC test as it does not attempt to model the dynamics of the stage. The CRNN method, however, is able to automatically find parameters governing the dynamics of stage 4, removing the need for approximations such as the cumulative enthalpy method.

\subsection{21700 battery cell simulation}

\begin{figure}[h!]
\centering
\begin{subfigure}{0.4\textwidth}
    \includegraphics[width=\textwidth]{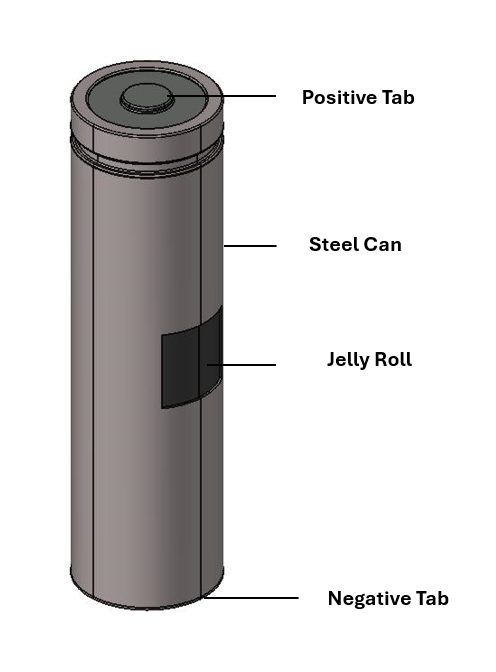}
    \caption{}
    \label{fig:cell_diagram}
\end{subfigure}
\hfill
\begin{subfigure}{0.48\textwidth}
    \centering
    \includegraphics[width=0.5\textwidth]{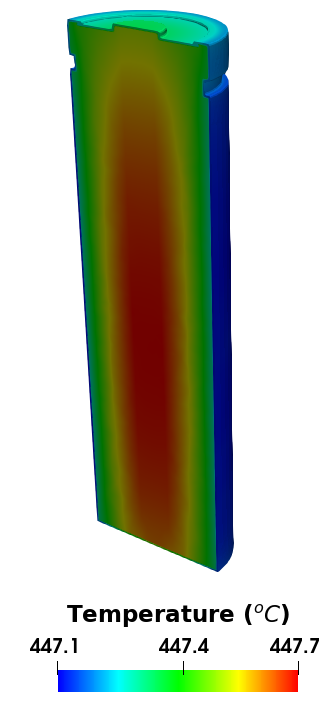}
    \caption{}
    \label{fig:heated_cell}
\end{subfigure}

\caption{\centering (\subref{fig:cell_diagram}) Geometry of cell showing the can, tabs and jellyroll 
(\subref{fig:heated_cell}) Simulated  temperature distribution during an ARC test}
\label{fig:3D_sim}
\end{figure}

To demonstrate the fitted parameters in a realistic scenario, 3D simulations of a single 21700 cell were conducted under ARC and oven test conditions. Figure (\ref{fig:cell_diagram}) depicts the battery geometry. Thermal runaway was modeled using the modified energy equation, which includes an additional source term \( S_{TR} \), given by

\begin{equation}
    \label{eqn:3D_PDE}
    \rho c_{p}\frac{\partial T}{\partial t}= \nabla\cdot(\kappa\nabla T) + S_{TR},
\end{equation}

where is $\rho$ the density, $c_p$ is the heat capacity, and $\kappa$ is the thermal conductivity. $S_{TR}$ is then given by the fitting values obtained from the 0D model, that is

\begin{equation}
    \label{eqn:STR}
    S_{TR}=\sum_{i=1}^{N}{h_i\frac{d\alpha_i}{dt}},
\end{equation}

Where $h_i$ represents the enthalpy of different battery components. $S_{TR}$ is active only within the jellyroll of the cell and is modeled using a four-stage thermal runaway model and the parameters obtained in Section \ref{sec:4_stage_results}. The governing energy equation includes boundary conditions for radiation and natural convection. In the ARC test, the ambient temperature is determined by the temperature recorded inside the chamber. For the oven test, the temperature reference is a single value representing the oven temperature. While more accurate simulations could incorporate natural convection volume surrounding the cell (e.g. the ARC chamber), these were simplified in the current simulations. Simulations were performed in the multi-purpose CFD software Altair\textsuperscript{\textregistered} AcuSolve\textsuperscript{\textregistered}. The governing Equation (\ref{eqn:3D_PDE}) is discretized using a Galerkin least squares stabilized finite element method. The battery geometry consisted of three main components: the jellyroll; the aluminium positive and negative tabs/terminals; and a steel can. A summary of the material properties for each of these components is provided in Table \ref{tab:3d_sim_physical_properties}. The jellyroll, an effective material representation of the true wound geometry, has cylindrical anisotropic properties for thermal conductivity; the through-plane conductivity is significantly lower than in-plane conductivity.

\begin{table}[]
    \centering
    \
\begin{tabular}{cccc} \toprule
    Component & Density (\(kg/m^3\)) & Specific Heat (\(J/kgK\)) & Conductivity (\(W/mK\))  \\ \midrule
    
     \textbf{Jelly Roll} & 161 & 191 & (\(k_{r},k_{\theta},k_{h}\))=(0.3,28,28)  \\  \midrule

     \textbf{Steel Can} & 7917 & 460 & 14  \\ \midrule

     \textbf{Tabs} & 2770 & 986 & 175  \\ 
     \bottomrule
\end{tabular}
\caption{Properties of cell components described in Figure \ref{fig:cell_diagram}}
    \label{tab:3d_sim_physical_properties}
\end{table}

\begin{figure}[h!]
\centering
\begin{subfigure}{0.48\textwidth}
    \includegraphics[width=\textwidth]{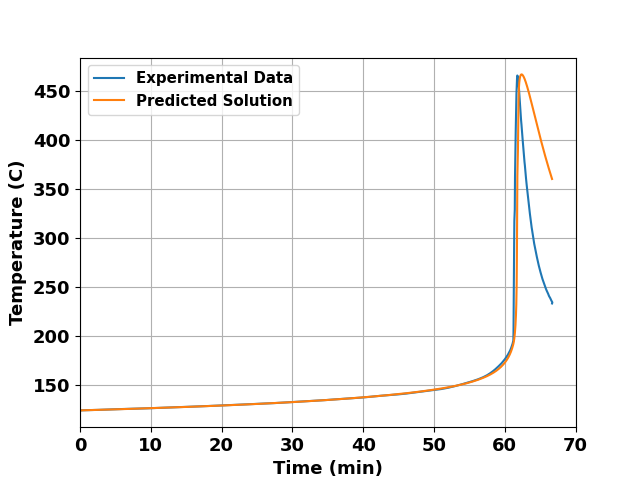}
    \caption{}
    \label{fig:4_stage_acusolve_ARC}
\end{subfigure}
\hfill
\begin{subfigure}{0.48\textwidth}
    \includegraphics[width=\textwidth]{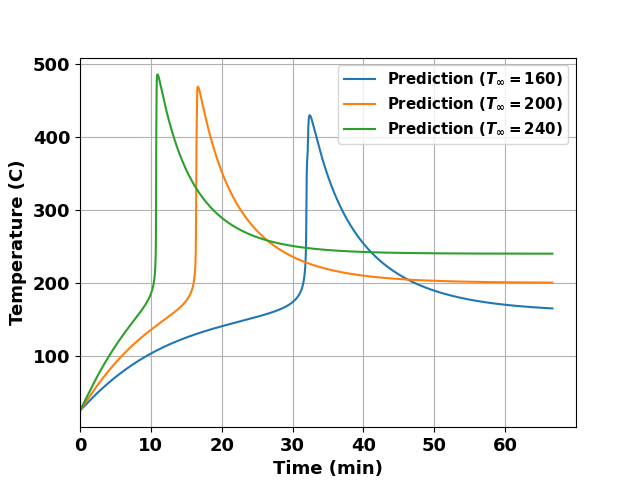}
    \caption{}
    \label{fig:4_stage_acusolve_oven}
\end{subfigure}
\hfill
\caption{\centering (\subref{fig:4_stage_acusolve_ARC}) Simulating the ARC test in 3D.  (\subref{fig:4_stage_acusolve_oven}) Simulating oven tests in 3D at various oven temperatures, observing an expected trend in temperature evolution.}
\label{fig:3D_acusolve_results}
\end{figure}

Figure (\ref{fig:heated_cell}) shows the temperature distribution in the cell during thermal runaway, highlighting the core of the battery heating up as the components of the jellyroll decompose. Figure (\ref{fig:4_stage_acusolve_ARC}) displays the mean temperature history from a 3D ARC test simulation, compared to experimental temperature measurements. The model accurately predicts the occurrence and peak temperature of thermal runaway, although some deviation afterwards was observed due to differing experimental and simulation conditions.

Oven test results in Figure (\ref{fig:4_stage_acusolve_oven}) for temperatures of 160$^o C$, 200$^o C$, and 240$^o C$ show expected trends: higher oven temperatures accelerate thermal runaway and result in higher peak temperatures. These findings demonstrate that the thermal runaway model, derived using parameters from CRNN, effectively predicts thermal behaviour in 3D simulations under varying physical and environmental conditions.

\section{Conclusion}

This work explored using Chemical Reaction Neural Networks (CRNNs) to perform kinetic parameter fitting of multi-stage thermal runaway models governed by the Arrhenius equations to ARC data. The issues with using linearization to obtain the kinetic parameters were discussed and demonstrated, and the linear fit results were improved using a CRNN for both two-stage and four-stage models. The parameters obtained showed much better agreement with the experimental data. The four-stage fit was also used to simulate ARC and oven tests in 3D, with results showing agreement with experiments. This further demonstrated the applicability of the derived thermal runaway models for simulations involving batteries or battery packs, where thermal runaway is a major safety concern and hence is a crucial design factor. The proposed method is flexible and can generalize to several stages, cell types, and stoichiometries, representing a step toward enhanced battery pack safety through simulation-driven design.

\section{CRediT Authorship Statement}

\hspace{5mm} \textbf{Saakaar Bhatnagar}:
Conceptualization, Formal Analysis, Methodology, Software, Visualization, Writing- Original Draft

\textbf{Andrew Comerford}: 
Conceptualization, Project Administration, Supervision, Validation, Visualization, Writing- Review and Editing

\textbf{Zelu Xu}: 
Conceptualization, Formal Analysis, Validation

\textbf{Davide Berti Polato}:  Data Curation, Investigation, Resources

\textbf{Araz Banaeizadeh}:  Project Administration, Supervision, Writing- Review and Editing 

\textbf{Alessandro Ferraris}: Data Curation, Investigation, Resources

\section{Funding Sources}

This research received no specific grant from funding agencies in the public, commercial, or not-for-profit sectors.

\bibliographystyle{unsrtnat}
\bibliography{bib}

\end{document}